\documentclass[11pt]{article}

\def\T{{\scriptstyle T}}
\def\bfu{{\bf u}}
\def\bfv{{\bf v}}

\def\bfy{{\bf y}}
\def\bft{{\bf t}}
\def\bfk{{\bf k}}
\def\bfn{{\bf n}}
\def\bfb{{\bf b}}
\def\bfC{{\bf C}}
\def\bfL{{\bf L}}
\def\bfK{{\bf K}}
\def\bfW{{\bf W}}

\input psfig.tex

\begin{document}

\fontsize{11}{14.5pt}\selectfont

\begin{center}

{\small Technical Report No.\ 9702,  
 Department of Statistics, University of Toronto}

\vspace*{0.5in}

{\Large \bf Monte Carlo Implementation of Gaussian Process Models for \\[6pt]
             Bayesian Regression and Classification} \\[16pt]

{\large Radford M. Neal}\\[1pt]
 Department of Statistics and Department of Computer Science \\
 University of Toronto, Toronto, Ontario, Canada \\
 \texttt{radford@stat.utoronto.ca}\\[10pt]

20 January 1997
\end{center}

\vspace{8pt} 

\noindent \textbf{Abstract.} Gaussian processes are a natural way of
defining prior distributions over functions of one or more input
variables.  In a simple nonparametric regression problem, where such a
function gives the mean of a Gaussian distribution for an observed
response, a Gaussian process model can easily be implemented using
matrix computations that are feasible for datasets of up to about a
thousand cases.  Hyperparameters that define the covariance function
of the Gaussian process can be sampled using Markov chain methods.
Regression models where the noise has a $t$~distribution and logistic
or probit models for classification applications can be implemented by
sampling as well for latent values underlying the observations.
Software is now available that implements these methods using
covariance functions with hierarchical parameterizations.  Models
defined in this way can discover high-level properties of the data,
such as which inputs are relevant to predicting the response.

\section{Introduction}\label{sec-intro}\vspace*{-10pt}

A nonparametric Bayesian regression model must be based on a prior
distribution over the infinite-dimensional space of possible
regression functions.  It has been known for many years that such
priors over functions can be defined using Gaussian processes (O'Hagan
1978), and essentially the same model has long been used in spatial
statistics under the name of ``kriging''.  Gaussian processes seem to
have been largely ignored as general-purpose regression models,
however, apart from the special case of smoothing splines (Wahba
1978), and some applications to modeling noise-free data from computer
experiments (eg, Sack, Welch, Mitchell, and Wynn 1989).  Recently, I
have shown that many Bayesian regression models based on neural
networks converge to Gaussian processes in the limit of an infinite
network (Neal 1996).  This has motivated examination of Gaussian
process models for the high-dimensional applications to which neural
networks are typically applied (Williams and Rasmussen 1996).  The
empirical work of Rasmussen (1996) has demonstrated that Gaussian
process models have better predictive performance than several other
nonparametric regression methods over a range of tasks with varying
characteristics.  The conceptual simplicity, flexibility, and good
performance of Gaussian process models should make them very
attractive for a wide range of problems.

One reason for the previous neglect of Gaussian process regression may
be that in a straightforward implementation it involves matrix
operations whose time requirements grow as the cube of the number of
cases, and whose space requirements grow as the square of the number
of cases.  Twenty years ago, this may have limited use of such models
to datasets with less than about a hundred cases, but with modern
computers, it is feasible to apply Gaussian process models to datasets
with a thousand or more cases.  It may also be possible to reduce the
time requirements using more sophisticated algorithms (Gibbs and
MacKay 1997a).

The characteristics of a Gaussian process model can easily be
controlled by writing the covariance function in terms of
``hyperparameters''.  One approach to adapting these hyperparameters
to the observed data is to estimate them by maximum likelihood (or
maximum penalized likelihood), as has long been done in the context of
spatial statistics (eg, Mardia and Marshall 1984).  In a fully Bayesian
approach, the hyperparameters are given prior distributions.
Predictions are then made by averaging over the posterior distribution for
the hyperparameters, which can be done using Markov chain Monte Carlo
methods.  These two approaches often give similar results (Williams
and Rasmussen 1996, Rasmussen 1996), but the fully Bayesian approach
may be more robust when the models are elaborate.

Applying Gaussian process models to classification problems presents
new computational problems, since the joint distribution of all
quantities is no longer Gaussian.  Approximate methods of Bayesian
inference for such models have been proposed by Barber and Williams
(1997) and by Gibbs and MacKay (1997b).  A general approach to exactly
handling classification and other generalized models (eg, for a
Poisson response) is to use a Markov chain Monte Carlo scheme in which
unobserved ``latent values'' associated with each case are explicitly
represented.  This paper applies this approach to classification using
logistic or probit models, and to regression models in which the noise
follows a $t$~distribution.

I have written software in C for Unix systems that implements Gaussian
process methods for regression and classification, within the same
framework as is used by my Bayesian neural network software.  This
software is is freely available for research and educational
use.\footnote{Follow the links from my home page, at
\texttt{http://www.cs.utoronto.ca/$\sim$radford/}.  The version
described here is that of 1997-01-18.} The covariance functions
supported may consist of several parts, and may be specified in terms
of hyperparameters, as described in detail in Section~\ref{sec-cov}.
These covariance functions provide functionality similar to that of
the neural network models.  The software implements full Bayesian
inference for these hierarchical models using matrix computations and
Markov chain sampling methods, as described in Sections~\ref{sec-mat}
and~\ref{sec-mc}.  In Sections~\ref{sec-exc} and~\ref{sec-exo}, I
demonstrate the use of the software on a three-way classification
problem, using a model that can identify which of the inputs are
relevant to predicting the class, and on a regression problem with
outliers.  I conclude by discussing some areas for future research.
First, however, I will introduce in more detail the idea of Bayesian
modeling using Gaussian processes.

\section{Regression and classification using Gaussian 
processes}\label{sec-gp}\vspace*{-10pt}

Assume we have observed data for $n$ cases,
$(x^{(1)},t^{(1)}),\,(x^{(2)},t^{(2)}),\,\ldots,\,(x^{(n)},t^{(n)})$,
in which $x^{(i)} = x^{(i)}_1,\ldots,x^{(i)}_p$ is the vector of $p$
``inputs'' (predictors) for case $i$ and $t^{(i)}$ is the associated
``target'' (response).  Our primary purpose is to predict the target,
$t^{(n+1)}$, for a new case where we have observed only the inputs,
$x^{(n+1)}$.  (We might sometimes be interested in interpretation as
well, but there is no point in interpreting a model that has failed to
capture the regularities that would support good predictive
performance.)  For a regression problem, the targets will be
real-valued; for a classification problem, the targets will be from
some finite set of class labels, which we will take to be
$\{0,\ldots,K\!-\!1\}$.  It will sometimes be convenient to represent
the distributions of the targets, $t^{(i)}$, in terms of 
unobserved ``latent values'', $y^{(i)}$, associated with each case.

Bayesian regression and classification models are usually formulated
in terms of a prior distribution for a set of unknown model
parameters, from which a posterior distribution for the parameters is
derived, and generally exhibited explicitly.  If our focus is on
prediction for a future case, however, the final result is a
predictive distribution for a new target value, $t^{(n+1)}$, that is
obtained by integrating over the unknown parameters.  This predictive
distribution can therefore be expressed directly in terms of the
inputs for the new case, $x^{(n+1)}$, and the inputs and targets for
the $n$ observed cases, without any mention of the model parameters.
What is more, rather than expressing our prior knowledge in terms of a
prior for the parameters, we can instead integrate over the parameters
to obtain a prior distribution for the targets in any set of cases.  A
predictive distribution for an unknown target can then be obtained by
conditioning on the known targets.  These operations are most easily
carried out if all the distributions are Gaussian.  Fortunately,
Gaussian processes are flexible enough to represent a wide variety of
interesting regression models, many of which would have an infinite
number of parameters if formulated in more conventional fashion.

Before discussing such nonparametric models, however, it may help
to see how the scheme works for a simple linear regression model,
which can be written as
\begin{eqnarray}
  t^{(i)} & = & \alpha\ +\ \sum_{u=1}^p x^{(i)}_u \beta_u\ +\ \epsilon^{(i)}
\end{eqnarray}
where $\epsilon^{(i)}$ is the Gaussian ``noise'' for case $i$, assumed to be 
independent from case to case, and to have mean zero and variance 
$\sigma^2_{\epsilon}$.
For the moment, we will assume that $\sigma^2_{\epsilon}$ is known, but that
$\alpha$ and the $\beta_u$ are unknown.  

Let us give $\alpha$ and the $\beta_u$ independent Gaussian priors
with means of zero and variances $\sigma^2_{\alpha}$ and
$\sigma^2_u$.  For any set of cases with fixed inputs,
$x^{(1)},\, x^{(2)},\,\ldots$, this prior distribution for parameters
implies a prior distribution for the associated target values,
$t^{(1)},\, t^{(2)},\,\ldots$, which will be multivariate Gaussian,
with mean zero, and with covariances given by
\begin{eqnarray}
  \mbox{Cov}\left[t^{(i)},\,t^{(j)}\right] & = &
  E\left[\left(\alpha\ +\ \sum_{u=1}^p x^{(i)}_u \beta_u\ +\ 
      \epsilon^{(i)}\right)
   \left(\alpha\ +\ \sum_{u=1}^p x^{(j)}_u \beta_u\ +\ \epsilon_j\right)
  \right] \\[4pt]
  & = &
  \sigma^2_{\alpha}\ +\ \sum_{u=1}^p x^{(i)}_u x^{(j)}_u \sigma^2_u
   \ +\ \delta_{ij} \sigma^2_{\epsilon}\label{eq-lin-cov}
\end{eqnarray}
where $\delta_{ij}$ is one if $i=j$ and zero otherwise.  This mean and
covariance function are sufficient to define a ``Gaussian process''
giving a distribution over possible relationships between the inputs
and the target. (Strictly speaking, one might wish to confine the term
``Gaussian process'' to distributions over functions from the inputs
to the target.  The relationship above is not functional, since (due
to noise) $t^{(i)}$ may differ from $t^{(j)}$ even if $x^{(i)}$ is
identical to $x^{(j)}$.  The looser usage is convenient here,
however.)

Suppose now that we know the inputs, $x^{(1)},\ldots,x^{(n)}$, for $n$ 
observed cases, as well as $x^{(n+1)}$, the inputs in a case for which we
wish to predict the target.  We can use equation~(\ref{eq-lin-cov})
to compute the $n\!+\!1$ by $n\!+\!1$ covariance matrix of the 
associated targets, $t^{(1)},\ldots,t^{(n)},t^{(n+1)}$.  Together 
with the assumption that the means are zero, these covariances
define a Gaussian joint distribution for the targets in the observed
and unobserved cases. We can condition on the known targets to obtain
the predictive distribution for $t^{(n+1)}$ given $t^{(1)},\ldots,t^{(n)}$.
Well-known results (eg, von Mises 1964, Section 9) show that this predictive 
distribution is Gaussian, with mean and variance given by
\begin{eqnarray}
  E\left[t^{(n+1)}\,\Big|\,t^{(1)},\ldots,t^{(n)}\right] & = &  
  \bfk^\T\, \bfC^{-1}\, \bft
  \label{eq-predm}\\[4pt]
  \mbox{Var}\left[t^{(n+1)}\,\Big|\,t^{(1)},\ldots,t^{(n)}\right] & = & 
  v\ -\ \bfk^\T\, \bfC^{-1}\, \bfk
\label{eq-predv}\end{eqnarray}
where $\bfC$ is the $n$ by $n$ covariance matrix for the targets in
the observed cases (from equation~(\ref{eq-lin-cov})), $\bft =
[t^{(1)}\,\cdots\,t^{(n)}]^\T$ is the vector of known target values in
these cases, $\bfk$ is the vector of covariances between $t^{(n+1)}$
and the $n$ known targets, and $v$ is the prior variance of
$t^{(n+1)}$ (ie, $\mbox{Cov}[t^{(n+1)},\,t^{(n+1)}]$ from 
equation~(\ref{eq-lin-cov})).

In practice, our prior knowledge will usually not be sufficient to fix
appropriate values for the ``hyperparameters'' that define the
covariance ($\sigma_{\epsilon}$, $\sigma_{\alpha}$, and the $\sigma_u$
for the simple model of equation~(\ref{eq-lin-cov})).  We will therefore 
give prior distributions to the hyperparameters, and base predictions on a
sample of values from their posterior distribution.  Sampling from the
posterior distribution requires computation of the log likelihood
based on the $n$ observed cases, which is\vspace{-4pt}
\begin{eqnarray}
  L & = & -\ {n\over2} \log(2\pi)\ -\ {1\over2}\,\log \det \bfC
          \ -\ {1\over2}\, \bft^\T\, \bfC^{-1}\, \bft
\label{eq-loglike}\end{eqnarray}
The derivatives of $L$ can also be computed, and used when sampling,
as described in Sections~\ref{sec-mat} and~\ref{sec-mc}.

This procedure is unnecessarily expensive for the simple regression
model just discussed, which is better handled by more standard
computational procedures.  However, the Gaussian process procedure can 
handle more interesting models by simply using a different covariance function
than that of equation~(\ref{eq-lin-cov}).  For example, a regression
model based on arbitrary smooth functions can be obtained using the
covariance function
\begin{eqnarray}
  \mbox{Cov}\left[t^{(i)},\,t^{(j)}\right] & = & \eta^2 
   \exp\left(- \sum_{u=1}^p\,\rho^2_u\,(x^{(i)}_u - x^{(j)}_u)^2 \right)
   \ +\ \delta_{ij} \sigma^2_{\epsilon}
\end{eqnarray}
Here, $\eta$ and the $\rho_u$ are hyperparameters, which would usually
be given some prior distribution rather than being fixed. Other possibilities 
for the covariance function are discussed in Section~\ref{sec-cov}.

Regression models with non-Gaussian noise and models for
classification problems, where the targets are from the set
$\{0,\ldots,K\!-\!1\}$, can be defined in terms of a Gaussian process
model for ``latent values'' associated with each case.  These latent
values are used to define a distribution for the target in a case.

For example, a logistic model for binary targets can be defined in
terms of latent values $y^{(i)}$ by letting the distribution for the 
target in case $i$ be given by
\begin{eqnarray}
  P(t^{(i)}=1) & = & \left[1+\exp\left(-y^{(i)}\right)\right]^{-1}
\label{eq-logistic}\end{eqnarray}
The latent values are given some Gaussian process prior, such as
\begin{eqnarray}
  \mbox{Cov}\left[y^{(i)},\,y^{(j)}\right] & = & \eta^2 
   \exp\left(- \sum_{u=1}^p\,\rho^2_u\,(x^{(i)}_u - x^{(j)}_u)^2 \right)
  \label{eq-nojit}
\end{eqnarray}
This covariance function gives a model in which the probability of
the target being 1 varies smoothly as a function of the inputs.

When there are three or more classes, an analogous model can be
defined using $K$ latent values for each case, $y^{(i)}_0,\ldots,y^{(i)}_{K-1}$,
which define class probabilities as follows:
\begin{eqnarray}
  P(t^{(i)}=k) & = & \exp\left(-y^{(i)}_k\right) \Big/\,
                \sum_{k^{\prime}=0}^{K-1}\exp\left(-y^{(i)}_{k^{\prime}}\right)
\label{eq-softmax}
\end{eqnarray}
The $K$ latent values can be given independent Gaussian process priors.
(This representation is redundant, but removing the redundancy by forcing
one of these latent values to always be zero would introduce an arbitrary 
asymmetry into the prior.)

For computational reasons, the covariance function of
equation~(\ref{eq-nojit}) must usually be modified by the addition of
at least a small amount of ``jitter'', as follows:
\begin{eqnarray}
  \mbox{Cov}\left[y^{(i)},\,y^{(j)}\right] & = & \eta^2 
   \exp\left(- \sum_{u=1}^p\,\rho^2_u\,(x^{(i)}_u - x^{(j)}_u)^2 \right)
   \ +\ \delta_{ij} J^2
\label{eq-withjit}\end{eqnarray}
Here, $J$ gives the amount of jitter, which is similar to the noise
in a regression model.  Including a small amount of jitter (eg, $J\!=\!0.1$)
makes the matrix computations better conditioned, and improves the 
efficiency of sampling, while having only a small effect on the 
model.

The effect of a probit model can be produced by using a larger amount of 
jitter.  A probit model for binary targets could be defined directly
in terms of latent values, $z^{(i)}$, having a covariance function 
without jitter, as follows:
\begin{eqnarray}
  P(t^{(i)}=1) & = & \Phi\!\left(z^{(i)}\right)
\label{eq-probit1}\end{eqnarray}
where $\Phi$ is the standard Gaussian cumulative distribution function.
This formulation of the probit model can be mimicked using latent
values, $y^{(i)}$, whose covariance function includes a jitter term
(as in equation~(\ref{eq-withjit})).  When $J\!=\!1$, the $y^{(i)}$ can 
be regarded as sums of jitter-free latent variables, $z^{(i)}$, and 
independent jitter of variance one.  A probit model can then be obtained 
using\vspace{-5pt}
\begin{eqnarray}
  P(t^{(i)}=1\ |\ y^{(i)}) & = & \Theta\!\left(y^{(i)}\right)
\label{eq-probit2}\end{eqnarray}
where $\Theta(y)=[\,0\ \mbox{if $y<0$};\ 1\ \mbox{if $y>=0$}\,]$. Integrating
over the jitter in $y^{(i)}$ gives the effect of equation~(\ref{eq-probit1}).
Finally, scaling up the magnitude of both the jitter and non-jitter parts of 
the covariance (eg, so that $J=10$) will leave the effect of 
equation~(\ref{eq-probit2}) unchanged, at which point the threshold function
can be replaced by the logistic function of equation~(\ref{eq-logistic}),
since the magnitude of $y^{(i)}$ will usually be large enough that the value
of the logistic function will be close to zero or one.\footnote{It would
be possible for the software to allow the option of using the $\Phi$ or 
$\Theta$ functions instead of the logistic, thereby allowing a probit model
to be implemented exactly, but this is not done at the moment.  It might also
be possible to allow the logistic to be replaced by another function that
produces the exact logistic model when some finite amount of jitter is used,
but this has not been investigated in detail either.}

If the covariance function used allows the latent values to be any
function of the inputs (plus jitter), the same class probabilities
will be representable using either a logistic or a probit model.  The
two sorts of models would differ only in the exact prior over class
probability functions that they embody.  It is not yet clear which of
the two models will be better in typical situations.  It is also
possible to make the amount of jitter be a hyperparameter, allowing
the data to determine which of the two models is more appropriate, or
to select an intermediate model.

Latent values can also be used to define regression models with
non-Gaussian noise, with the latent value being the noise-free value
of the regression function.  (In practice, it is usually necessary to
include a small amount of jitter in the covariance function for the
latent values, which has the effect of introducing some minimum amount
of Gaussian noise.)  A $t$~distribution for the noise is particularly
convenient, since it can be expressed in terms of a Gaussian noise
model in which the noise variances for the cases are independently
drawn from an inverse gamma distribution.  In the implementation of
this model, these case-by-case noise variances are explicitly
represented, and sampled.  The latent values are needed to sample for
the noise variances, but can be discarded once used for this purpose.

\section{Covariance functions and their 
hyperparameters}\label{sec-cov}\vspace*{-10pt}

A wide variety of covariance functions can be used in the Gaussian
process framework, subject to the requirement that a valid covariance
function must result in a positive semidefinite covariance matrix for
the targets in a set of any number of cases, in which the inputs take
on any possible values.  In a Bayesian model, the covariance function
will usually depend on various ``hyperparameters'', which are
themselves given prior distributions.  Such hyperparameters can
control the amount of noise in a regression model, the scale of
variation in the regression function, the degree to which various
input variables are relevant, and the magnitudes of different additive
components of a model.  The posterior distribution of these
hyperparameters will be concentrated on values that are appropriate
for the data that was actually observed.

Characterizing the set of valid covariance functions is not trivial,
as seen by the extensive discussions in the book by Yaglom (1987).
One way to construct a variety of covariance functions is by adding
and multiplying together other covariance functions, since the
element-by-element sum or product of any two symmetric, positive
semidefinite matrices is also symmetric and positive semidefinite
(Horn and Johnson 1985, 7.1.3 and 7.5.3).  Sums of covariance
functions are useful in defining models with an additive structure,
since the covariance function for a sum of independent Gaussian
processes is simply the sum of their separate covariance functions.
Products of covariance functions are useful in defining a covariance
function for cases with multidimensional inputs in terms of covariance
functions for single inputs.

The current software supports covariance functions that are the sum of
one or more terms of the following types:\vspace{-9pt}\begin{enumerate}
\item[1)] A constant part, which is the same for any pair of cases, 
          regardless of the inputs in those cases.  This adds a constant
          component to the regression function (or to the latent values for
          a classification model), with the prior for the value of this 
          constant component having the variance given by this constant 
          term in the covariance function.
\item[2)] A linear part, which for the covariance between cases $i$ and $j$
          has the form
          \begin{eqnarray}
             \sum\limits_{u=1}^p x^{(i)}_u x^{(j)}_u \sigma^2_u
          \end{eqnarray}
          This produces a linear function of the inputs, as seen in
          Section~\ref{sec-gp}, or adds a linear component to the 
          function, if there are other terms in the covariance as well.
\item[3)] A jitter part, which is zero for different cases, and a constant
          for the covariance of a case with itself.  Jitter is used
          to improve the conditioning of the matrix computations, or to
          produce the effect of a probit classification model.  The
          noise in a regression model is similar, but is treated separately
          in this implementation (jitter affects the latent values, and through
          them the targets; noise affects only the targets).
\item[4)] Any number of exponential parts, each of which, for
          the covariance between cases $i$ and $j$, has the form\vspace{-6pt}
          \begin{eqnarray}
             \eta^2 \prod_{u=1}^p
             \exp\left(-
                \left(\,\rho_u\left|x^{(i)}_u - x^{(j)}_u\right|\,\right)^R
             \right)
          \label{eq-exp-cov}\end{eqnarray}          
          If there are several exponential parts, they may use different
          values of $R$, $\eta$, and the~$\rho_u$. For
          the covariance function to be positive definite, $R$ must 
          be in the range 0 to~2.  The default value of $R\!=\!2$ produces 
          a function (or additive component of a function) that is infinitely 
          differentiable, but not constrained to be of any particular
          form.\vspace{-9pt}
\end{enumerate}
The parameters of these terms in the covariance function may be fixed,
or they may be treated as hyperparameters, with given prior distributions,
except that the power, $R$, for an exponential part must currently
be fixed.  

\begin{figure}[p]

\vspace*{2.5in}

\includegraphics{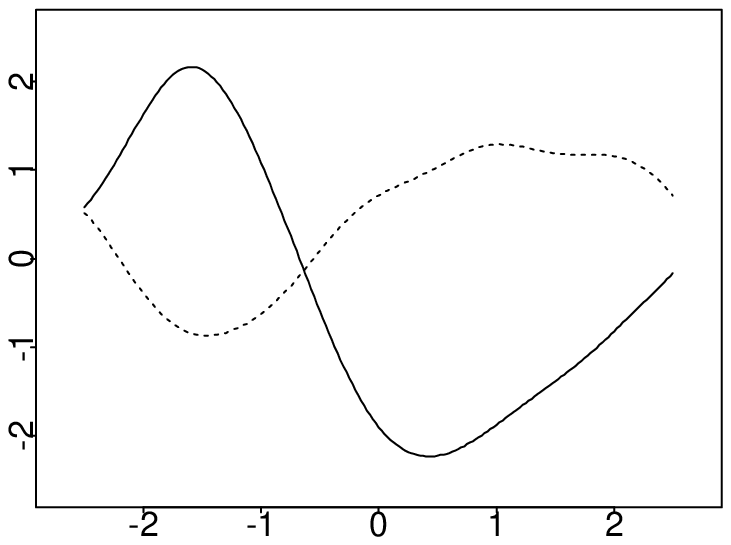}
\includegraphics{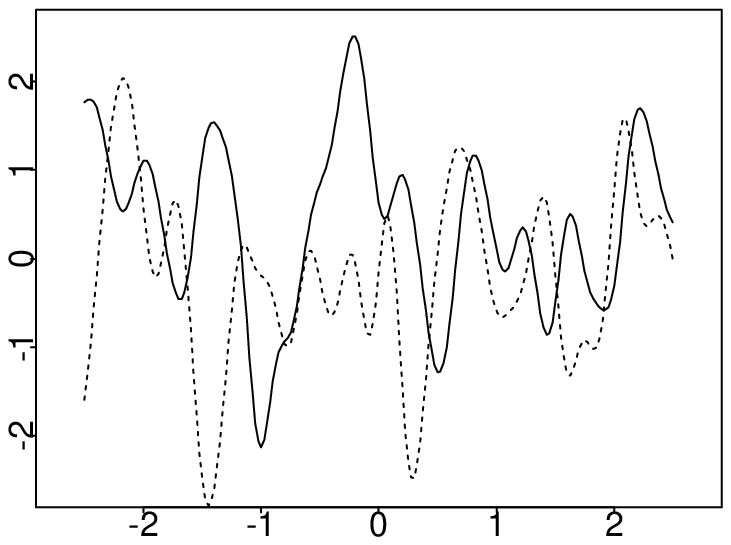}

\vspace*{-0.15in}

$\ \ \ \mbox{Cov}\left[t^{(i)},\,t^{(j)}\right] = 
  \exp\Big(\!-\left(x^{(i)}-x^{(j)}\right)^2\Big)$
\hfill 
$\mbox{Cov}\left[t^{(i)},\,t^{(j)}\right] = 
  \exp\Big(\!-5^2\left(x^{(i)}-x^{(j)}\right)^2\Big)\ $

\vspace*{2.8in}

\includegraphics{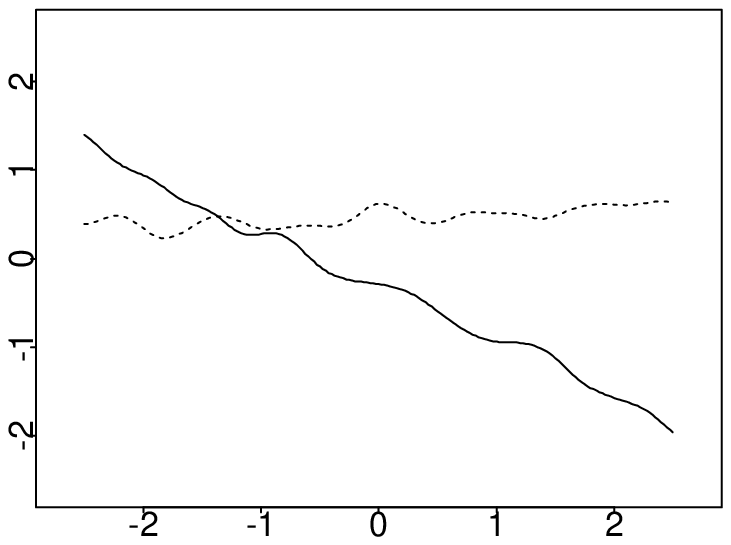}
\includegraphics{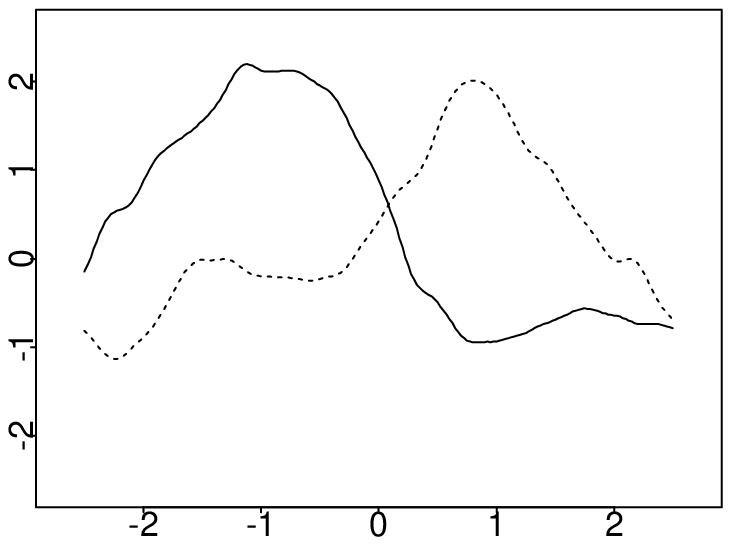}

\vspace*{-0.15in}

$\ \ \ \ \ \ \mbox{Cov}\left[t^{(i)},\,t^{(j)}\right]\ =\ 
   1\ +\ x^{(i)}x^{(j)}$
\hfill 
$\mbox{Cov}\left[t^{(i)},\,t^{(j)}\right]\ =\  
  \exp\Big(\!-\left(x^{(i)}-x^{(j)}\right)^2\Big)\ \ $

\vspace{5pt}

$\ \ \ \ \ \ \ \ \ \ \ \ \ \
 \ +\ 0.1^2\,\exp\Big(\!-3^2\left(x^{(i)}-x^{(j)}\right)^2\Big)\ $
\hfill 
$\ +\ 0.1^2\,\exp\Big(\!-5^2\left(x^{(i)}-x^{(j)}\right)^2\Big)\ \ $

\vspace*{0.2in}

\caption[]{Functions drawn from Gaussian processes with various
covariance functions.  Each of the graphs shows two functions
that were independently drawn from the Gaussian process with
mean zero and with the covariance function given below the 
graph.}\label{fig-prior}

\end{figure}

Some of the possible distributions over functions that can be obtained
using covariance functions of this form are illustrated in
Figure~\ref{fig-prior}.  (These are functions of a single input, so
the index $u$ is dropped).  The top left and top right each show
functions drawn randomly from a Gaussian process with a covariance
function consisting of a single exponential part.  The distance over
which the function varies by an amount comparable to its full range,
given by $1/\rho$, is smaller for the top-right than the top-left.
The bottom left shows functions generated using a covariance function
that is the sum of constant, linear, and exponential parts.  The
magnitude of the exponential part, given by $\eta$, is rather
small, so the functions depart only slightly from straight lines.  The
bottom right shows functions drawn from a prior whose covariance
function is the sum of two exponential parts, that produce variation
at different scales, and with different magnitudes.  The software can
produce such plots of randomly drawn functions in one and two
dimensions, using the Cholesky decomposition of the covariance matrix
for the targets over a grid of input points, as described in
Section~\ref{sec-mat}.

For problems with more than one input variable, the $\sigma_u$ and
$\rho_u$ parameters control the degree to which each input is relevant
to predicting the target.  If $\rho_u$ is close to zero, input $u$
will have little effect on the degree of covariance between cases (or
at least, little effect on the portion of the covariance due to the
exponential part in which the $\rho_u$ hyperparameter occurs).  Two
cases could therefore have high covariance even if they have greatly
different values for input $u$ --- ie, input $u$ is effectively
ignored.

In typical applications, the constant part and jitter part (if any) of
the covariance would be given fixed values, but the available prior
information would not be sufficient to fix the other hyperparameters,
which specify the magnitudes of the linear and exponential parts, and
the scales of variation and relevances of the various inputs.  The
standard deviation of the noise for a regression model would also
typically be unknown.  These hyperparameters should therefore usually
be given fairly vague prior distributions.  These distributions should
be proper, however, as using an improper prior will often produce an
improper posterior.

The priors for hyperparameters supported by the software all take 
the same form.  If $\theta$ is a hyperparameter (all of which take on
only positive values), the value of $\phi=\theta^{-2}$ can be given a
gamma prior with density 
\begin{eqnarray}
  p(\phi) & = & {(\alpha/2\omega)^{\alpha/2} \over \Gamma(\alpha/2)}\,
                \phi^{\alpha/2-1}\,\exp(-\phi\alpha/2\omega)
\end{eqnarray}
Here, $\alpha$ is a positive shape parameter, and $\omega$ is the
mean of $\phi$.  The software accepts prior specifications in
terms of $\alpha$ and $\omega^{-2}$ (whose units correspond
to those of the original hyperparameter, $\theta$).  Large values of
$\alpha$ produce priors for $\theta$ concentrated near $\omega^{-2}$, 
whereas small values of $\alpha$ produce vague priors.\footnote{There is 
an arbitrary aspect to this form of prior specification, since $\alpha$
controls not only how diffuse the prior is, but also its shape.  This
could be fixed by letting the gamma prior be for $\phi=\theta^r$, with
$r$ being any specified power.  The present scheme is analogous to
the priors used for neural network models, where $r=-2$ results in
a conjugate prior with some computational advantages.}

Single hyperparameters, such as $\eta$ in an exponential part, or the
noise standard deviation for a simple regression model, may either be
given a prior as described above, or be given a fixed value (equivalent
to letting $\alpha=\infty$).  Hyperparameters that come in groups,
such as the $\rho_u$ in an exponential part, or the $\sigma_u$ in a
linear part, can be given hierarchical priors, expressed in terms of a
higher-level hyperparameter associated with the group, which has no
direct effect on the covariance function, but which determines the
mean for the lower-level hyperparameters.  For example, the $\rho_u$
hyperparameters for an exponential part might be accompanied by a
higher-level hyperparameter, $\rho_*$. At the top level, $\phi_* =
\rho_*^{-2}$ could be given a gamma prior of the form 
\begin{eqnarray}
  p(\phi_*) & = & {(\alpha_0/2\omega)^{\alpha_0/2} \over \Gamma(\alpha_0/2)}\,
                \phi_*^{\alpha_0/2-1}\,\exp(-\phi_*\alpha_0/2\omega)
\end{eqnarray}
For a given value of $\rho_*$, the $\rho_u$ hyperparameters associated
with particular inputs are independent, with $\phi_u=\rho_u^{-2}$
having a gamma prior with mean $\phi_*$, as follows:
\begin{eqnarray}
  p(\phi_u\,|\,\phi_*) 
  & = & {(\alpha_1/2\phi_*)^{\alpha_1/2} \over \Gamma(\alpha_1/2)}\,
        \phi_u^{\alpha_1/2-1}\,\exp(-\phi_u\alpha_1/2\phi_*)
\end{eqnarray}
Note that the shape parameters for the two levels, $\alpha_0$ and $\alpha_1$,
can be different ($\alpha_1$ is the same for all inputs, $u$, however).
The top level of the hierarchy can be omitted (effectively, $\alpha_0=\infty$),
in which case the $\rho_u$ are independent.  The lower level of the 
hierarchy can also be omitted (effectively, $\alpha_1=\infty$), in which
case the $\rho_u$ are all equal to $\rho_*$.  Finally both levels can be omitted
($\alpha_0=\alpha_1=\infty$), in which case the $\rho_u$ have fixed values.

These hierarchical priors can also link together the $\sigma_u$ parameters in
the linear part of the covariance, as well as the noise standard deviations
for a regression model with more than one target.  However, at present
there is no way of linking hyperparameters of different types, nor of
linking hyperparameters pertaining to different parts of the
covariance function (eg, the $\rho_u$ for different exponential
parts).  When there is more than one target or latent value for each
case, the same hyperparameters are currently used for the independent
Gaussian processes that model the relationship of each value to the
inputs.  The only exception to this is that different noise standard
deviations are possible for regression models with more than one
target.

In contrast to the elaborate provisions for different covariance
functions, the software currently assumes that the mean function for
the Gaussian process is always zero.  This is appropriate for problems
where prior knowledge is vague.  Note that using a zero mean Gaussian
process does \emph{not} mean that we expect the actual regression
function to take on positive and negative values over equal parts of
its range.  If the covariance function has a large constant term, we
would not be surprised if the actual function were always positive, or
always negative (at least over the range of interest).  Using a mean
function of zero simply reflects our lack of prior knowledge as to
what the sign will turn out to be.  In practice, it will usually be
desirable to transform the targets so that their mean is approximately
zero, in order to eliminate any need for a large constant term in the
covariance.  Including a large constant term is undesirable because it
increases the round-off error in the matrix computations.

\section{Matrix computations}\label{sec-mat}\vspace*{-10pt}

Inferences regarding a Gaussian process model given particular values
for its hyperparameters can be performed using computations involving
the covariance matrix for the targets or latent values associated with
the observed cases.  If appropriate hyperparameter values are known
\emph{a priori}, these matrix computations are all that are needed to
make predictions for the targets in new cases.  In the more common
situation where the hyperparameters are unknown, such matrix
computations are used to support the Markov chain sampling methods
described in Section~\ref{sec-mc}, and then to make predictions using
the resulting sample of hyperparameter values (and latent values, if
required).

The central object in these computations is the $n$ by $n$ covariance
matrix of the latent values underlying the observed cases, or of the
target values themselves, for a regression model.  This covariance
matrix, which we will denote by $\bfC$, depends on the observed inputs
for these cases, and on the particular values of the hyperparameters,
both of which are considered fixed here.  The difficulty of
computations involving $\bfC$ is determined by its condition number
--- the ratio of its largest eigenvalue to its smallest eigenvalue.
If the condition number is large, round-off error in the matrix
computations may cause them to fail or to be highly inaccurate.  This
potential problem can be controlled by using covariance functions that
include ``jitter'' terms (see Section~\ref{sec-cov}), since the jitter
contributes additively to every eigenvalue of the matrix, reducing the
condition number.  When $\bfC$ is the covariance matrix for the
targets in a regression model, the noise variance has an equivalent
effect.  For most problems, it appears that the addition of a small
amount of jitter to the covariance will not seriously affect the
statistical properties of the model, and may even be desirable.
Accordingly, the software does not attempt to handle covariance
matrices that are very badly conditioned.

The present implementation is based on finding the Cholesky
decomposition of $\bfC$ --- that is, the lower-triangular matrix,
$\bfL$, for which $\bfC = \bfL \bfL^\T$.  The Cholesky decomposition
can be found by a simple algorithm (see, for example, Thisted 1988,
Section~3.3), which runs in time proportional to $n^3$.  Once the
Cholesky decomposition has been found, the determinant of $\bfC$ can
easily be computed as the square of the product of the diagonal
elements of $\bfL$.  (In practice, the log of the determinant is found
from the sum of the logs of the diagonal elements.) Another use of the
Cholesky decomposition is in generating latent or target values from
the prior.  Standard methods can be used to randomly generate a
vector, $\bfn$, composed of $n$ independent Gaussian variates with
mean zero and variance one.  One can then compute the vector $\bfL
\bfn$, which will have mean zero and covariance matrix $\bfL \bfL^\T =
\bfC$.  This procedure was used to produce the plots in
Figure~\ref{fig-prior}, using the covariance matrix for the targets
over a grid of input values, with the addition of an unnoticeable
amount of jitter.

The primary use of the Cholesky decomposition is in computing the
inverse of $\bfC$, which arises both in the predictive distribution
for a new case (equations~(\ref{eq-predm}) and~(\ref{eq-predv})) and
in the log likelihood (equation~(\ref{eq-loglike})). These
computations could be performed without explicitly finding the inverse
of $\bfC$, since $\bfC^{-1}\,\bfb$ can be found using the Cholesky
decomposition by first solving $\bfL \bfv = \bfb$ for $\bfv$ using
forward substitution, and then solving $\bfL^\T \bfu = \bfv$ for
$\bfu$ using backward substitution.  However, it is more convenient to
explicitly compute the inverse, since it will often be needed anyway
in order to compute derivatives of the log likelihood.  Computation of
$\bfC^{-1}$ is done by applying the procedure just described to
compute $\bfC^{-1}\,\bfb$ for the $n$ vectors $\bfb$ that are all zero
except for one element with the value one.  This takes time
proportional to $n^3$.

Once $\bfC^{-1}$ has been computed, we can prepare to make predictions
from a regression model by computing $\bfb = \bfC^{-1}\,\bft$, where
$\bft$ is the vector of targets in the training cases.  The mean of
the predictive distribution for the target in a test case can then be
found in time proportional to $n$.  We first compute the vector,
$\bfk$, of covariances between the targets in the test case and in the
$n$ training cases.  We then compute the predictive mean of
equation~(\ref{eq-predm}) as $\bfk^\T\, \bfb$.  This is the method
used in the present implementation.  An alternative is to solve $\bfL
\bfu = \bfk$ for $\bfu$ and to solve $\bfL \bfv = \bft$ for $\bfv$,
and then compute the predictive mean as $\bfu^\T \bfv$.  This is less
efficient, taking time proportional to $n^2$ for each test case, but
Gibbs and Mackay (1997a) report that it is more accurate when $\bfC$ is
poorly conditioned.  If we require the predictive variance as well as
the mean, we must compute $\bfk^\T\, \bfC^{-1}\, \bfk$, for use in
equation~(\ref{eq-predv}), which will take time proportional to $n^2$.

Predictions for classification models involve similar operations,
but focused on the latent value associated with a test case.  A vector
of latent values, $\bfy$, associated with the training cases must be
available.  The vector of covariances, $\bfk$, between these latent
values and the latent value in a test case can be computed, and a
predictive mean and variance for the latent value in the test case can
then be found as above.  A sample of values from this Gaussian predictive
distribution can easily be obtained, from which Monte Carlo estimates
for the class probabilities in the test case can be computed by simply
averaging the probabilities that are obtained by substituting the
latent values in this sample into equation~(\ref{eq-logistic})
or~(\ref{eq-softmax}).

One may also wish to sample from the joint posterior distribution of
the targets or latent values in a set of cases, either as part of some
other computation, or in order to plot regression functions drawn from
the posterior distribution.  Conditional on values for the hyperparameters,
for the latent variables associated with training cases (for a 
classification model), and for the case-by-case noise variances (for a
regression model with $t$-distributed noise), these distributions will be Gaussian, with means and covariances given by generalizations of 
equations~(\ref{eq-predm}) and~(\ref{eq-predv}).  In a regression model, 
for example, if $\bfy$ is the vector of latent values in a set of $m$ 
test cases, and $\bft$ is the vector of target values in $n$ training cases,
then\vspace*{-5pt}
\begin{eqnarray}
  E\left[\,\bfy\,|\,\bft\,\right] & = &  
  \bfK^\T\, \bfC^{-1}\, \bft
  \label{eq-predm-g}\\[4pt]
  \mbox{Cov}\left[\,\bfy\,|\,\bft\,\right] & = & 
  \bfW\ -\ \bfK^\T\, \bfC^{-1}\, \bfK
\label{eq-predv-g}\end{eqnarray}
where $\bfC$ is the $n$ by $n$ covariance matrix for the targets in 
training cases, $\bfW$ is the $m$ by $m$ covariance matrix for latent
values in the test cases, and $\bfK$ is the $n$ by $m$ matrix of
covariances between targets in training cases and latent values in
test cases.  Once these means and covariances have been computed,
a value for $\bfy$ can be generated using the Cholesky decomposition
of its covariance matrix, as described above in regard to sampling
from the prior.

The Markov chain methods used to sample from the posterior
distribution of the hyperparameters in a regression model require
computation of the log likelihood, $L$, of equation~(\ref{eq-loglike}).  
As seen above, this is easily done using the Cholesky decomposition of
$\bfC$.  For some of the Markov chain sampling methods, the derivatives 
of $L$ with respect to the various hyperparameters are also required.
The derivative of the log likelihood with respect to a hyperparameter
$\theta$ can be written as follows (Mardia and Marshall 1984):
\begin{eqnarray}
  {\partial L \over \partial \theta} & = & 
  - {1\over2}\,\mbox{tr} \left( \bfC^{-1}\,{\partial \bfC \over \partial \theta}
                         \right)
  \ +\ {1\over2}\, \bft^\T\, \bfC^{-1}\, {\partial \bfC \over \partial \theta}\,
                             \bfC^{-1}\, \bft
\label{eq-deriv-loglike}\end{eqnarray}
The trace of the product in the first term can be computed in time
proportional to $n^2$, assuming that $\bfC^{-1}$ has already been
computed.  The second term can also be computed in time proportional
to $n^2$, by first computing $\bfb = \bfC^{-1}\,\bft$ (which is
probably needed anyway, to compute $L$ itself), multiplying
this on the left by the matrix of derivatives, and finally multiplying 
the result by $\bfb^\T$.  Apart from the computation of $\bfb$, this
procedure must be repeated for each hyperparameter, and for each
regression target, if there is more than one. 

The Markov chain methods used for classification models require a
similar computation, but with the vector of targets, $\bft$, replaced
by the vector of current latent values, $\bfy$.

For large data sets, the time required for these computations is
dominated by that required to form the Cholesky decomposition of
$\bfC$, and to then compute $\bfC^{-1}$, for which the number of
operations required grows in proportion to $n^3$.  Indeed, on machines
with memory caches, the time required for these computations may grow
at a rate even faster than $n^3$, since larger matrices will not fit
in the fast cache.  The software attempts to reduce such cache effects
by whenever possible scanning matrices along rows rather than down
columns, but for large matrices the slowdown on our SGI machine can
still be substantial.

For small data sets (eg, 100 cases), the time required to compute the
derivatives of the log likelihood with respect to the hyperparameters
can dominate, even though this time grows only in proportion to $n^2$.
This may occur, for example, when there are many hyperparameters
controlling the relevance of many input variables, so that computing
the matrix of derivatives of the covariances takes a lot of time.
These computations can be sped up if the individual values for the
exponential parts of the covariances have been saved (as these appear
in the expressions for the derivatives).  The software does this when
$n$ is small enough that the memory required to do so is not too
large; when $n$ is larger, the other operations dominate anyway.

\section{Markov chain sampling}\label{sec-mc}\vspace*{-10pt}

The covariance functions for most Gaussian process models will contain
unknown hyperparameters, which must be integrated over in a fully
Bayesian treatment.  The number of hyperparameters will vary from
around three or four for a very simple regression model up to several
dozen or more for a model with many inputs, whose relevances are
individually controlled using hyperparameters such as the $\rho_u$ of
equation~(\ref{eq-exp-cov}).  Markov chain Monte Carlo methods (see
(Neal 1993) for a review) seem to be the only feasible approach to
performing these integrations, at least for the more complex models.
For classification models, latent values for each training case must
also be integrated over, and for regression models in which the noise
has a $t$~distribution, we must integrate over the case-by-case noise
variances.  These latent values and variances can be included in the
state of the Markov chain and sampled along with the hyperparameters.

Sampling from the posterior distribution of the hyperparameters is
facilitated by representing them in logarithmic form, as this makes
the sampling methods independent of the scale of the data.  The
widely-used method of Gibbs sampling cannot easily be applied to this
problem, since it seems difficult to sample from the conditional
distributions for one hyperparameter given values for the others (and
the latent values, if any).  The Metropolis algorithm could be used
with some simple proposal distribution, such as a Gaussian with
diagonal covariance matrix.  The software supports this option, along
with a variety of other Markov chain sampling methods.  However,
simple methods such as this explore the region of high probability by
an inefficient random walk.  It is probably better for most models to
use a method that can suppress these random walks (Neal 1993, 1996).

The most appropriate way to suppress random walks for this problem
seems to be to use the hybrid Monte Carlo method of Duane, Kennedy,
Pendleton, and Roweth (1987), or the variant of this method due to
Horowitz (1991).  I have employed the hybrid Monte Carlo method to do
Bayesian inference for neural network models (Neal 1996), and
Rasmussen (1996) has used it for Gaussian process regression.  Several
variants of the hybrid Monte Carlo method are supported by the Markov
chain modules that I use for both the neural network and the Gaussian
process software.  I will give only a brief, informal description of
the method here.  More details can be found elsewhere (Neal 1993,
1996; Rasmussen 1996).

The hybrid Monte Carlo method suppresses random walks by introducing
``momentum'' variables that are associated with the ``position''
variables that are the focus of interest.  For the Gaussian process
application, the position variables are the hyperparameters defining
the covariance function.  The state of the simulation evolves in the
same way as the position and momentum of a physical particle
travelling through a region of variable potential energy.  The
momentum causes the particle to continue in a consistent direction
until such time as a region of high energy (low probability) is
encountered.  This motion must be randomized a bit in order to ensure
that the correct distribution is sampled from, but not so much that
undesirable random walk behaviour results.

In practice, the differential equations that describe how the position
and momentum change through time are discretized, and the bias due to
discretization error is eliminated by accepting or rejecting the new
state in the Metropolis style.  The ``leapfrog'' discretization is
usually used.  In order to perform a leapfrog update, the derivatives
of the log of the posterior probability with respect to the
hyperparameters must be computed.  To decide whether to accept an
update (or sequence of updates), the log of the posterior probability
must be found (except for its normalizing constant).  The log
posterior probability is computed from the log of the prior
probabilities for the hyperparameters, which have the easily computed
gamma form, and the log likelihood, from equation~(\ref{eq-loglike}).
The derivatives are found by adding the derivative of the log prior,
which is easily computed, to the derivative of the log likelihood,
which is computed using equation~(\ref{eq-deriv-loglike}).  In the
original hybrid Monte Carlo method of Duane, \textit{et al.} (1987),
several leapfrog updates are done, after which a decision whether to
accept the result is made.  The momentum is also randomized at this
time.  A variation using ``windows'' of states (Neal 1994) can be used
to increase the acceptance probability.  In the variation due to
Horowitz (1991), an acceptance decision is made after each leapfrog
update, after which the momentum is only partially randomized. I refer
to this as hybrid Monte Carlo with ``persistence'' of the momentum.

For hybrid Monte Carlo to work well, appropriate ``stepsizes'' for the
leapfrog updates must be selected --- if too large a stepsize is used,
the acceptance rate will be very low, but if the stepsize is too
small, progress will be needlessly slow.  Different stepsizes can be
used for different hyperparameters; this is equivalent to rescaling
the hyperparameters (in their logarithmic form) using different scale
factors.  The software includes a heuristic procedure that
automatically selects a stepsize for each hyperparameter.  These
selections are based on estimates of the second derivatives of the log
posterior density with respect to the hyperparameters, which indicate
how large a change can be made to a hyperparameter without getting
into a region of low probability.  These automatically selected
stepsizes can be (and usually are) manually adjusted by multiplying
them all by some factor, which is chosen on the basis of preliminary
runs.  Accordingly, the real role of the heuristics is to set the
relative stepsizes for different hyperparameters.

The heuristics used at present are rather simple.  The stepsizes for
high-level hyperparameters are scaled down by the square root of the
number of low-level hyperparameters that they control.  This is in
accord with how one would expect the width of their posterior
distribution to scale.  Similarly, the stepsize for the noise variance
in a regression model is scaled down by the square root of the number
of training cases.  However, the stepsizes for the other
hyperparameters (eg, the $\eta$ and $\rho_u$ hyperparameters in an
exponential part of the covariance) are \textit{not} scaled on the
basis of the number of training cases.  Whether this is the right
thing to do depends on whether the posterior distribution for these
hyperparameters becomes more tightly concentrated as the number of
training cases increases.  I conjecture that these posterior
distributions are typically more concentrated than the prior, but that
they do not become more and more concentrated as the number of
training cases increases, except perhaps for the $\rho_u$ parameters
in an exponential part with $R<2$, for which the functions produced
are fractal.  Mardia and Marshall (1984) consider this problem in a
spatial statistics context, under the assumption that the range of the
input variables increases with the number of training cases, which I
presume is not the typical situation for regression and classification
problems.  If additional training cases instead provide denser
sampling within a fixed region, it seems that they can provide only a
limited amount of information about the hyperparameters, unless the
function modeled has a fractal nature, in which information is
repeated at all scales.

For a classification model, these hybrid Monte Carlo updates of the
hyperparameters use the likelihood based on the current latent values
associated with training cases, not on the targets directly.  These
hyperparameter updates must be interleaved with updates of the latent
values themselves, for which Gibbs sampling is presently used.  New
latent values are chosen for each case in a sequential
scan.\footnote{When there are several latent values for each case, it
makes no difference whether the inner loop of the scan is over cases
or over the several values for one case.}  These values are drawn from
the conditional distribution for such a latent value given the
observed target for that training case, and given the current values
of the hyperparameters and all the other latent values.  The density
for this conditional distribution is proportional to the product of
the likelihood given the target, from equation~(\ref{eq-logistic})
or~(\ref{eq-softmax}), and the Gaussian conditional density given the
other latent values.  The conditional density for $y^{(i)}$ given the
other latent values, all of which are collected in $\bfy$, is
proportional to $\exp(-{1\over2}\bfy^\T\bfC^{-1}\bfy)$, and can be
found in time proportional to $n$ if $\bfC^{-1}$ has already been
computed.  The final conditional density is log-concave, and hence can
efficiently be sampled from using the adaptive rejection method of
Gilks and Wild (1992).

Once $\bfC^{-1}$ has been computed, taking time proportional to $n^3$,
a complete Gibbs sampling scan takes time proportional only to $n^2$.
It therefore makes sense to perform quite a few Gibbs sampling scans
between each update of the hyperparameters, as this adds little to the
time requirements, and probably makes the Markov chain mix faster.

The software also supports regression models with $t$-distributed
noise, expressed as Gaussian noise with case-by-case variances drawn
from an inverse gamma distribution.  The Markov chain must then sample
somehow for the case-by-case noise variances, which are needed to
compute the covariances of the targets.  In one approach, case-by-case
latent values are maintained, and updated using Gibbs sampling, in a
manner analogous to that used for classification models.  Gibbs
sampling can then easily be done for the case-by-case noise variances
as well, based only on the hyperparameters controlling the noise
level, the latent values, and the targets.  The software also supports
a second approach, however, in which latent values are not kept around
permanently.  Instead, latent values are temporarily generated just
before the noise variances are updated, using
equations~(\ref{eq-predm-g}) and~(\ref{eq-predv-g}), and then
discarded after being used to generate new values for the noise
variances.

\section{Example:\ \ A three-way classification 
problem}\label{sec-exc}\vspace*{-10pt}

\begin{figure}[t]

\vspace*{40pt}
\hspace{10pt}\parbox{170pt}{
\caption[]{The 400 training cases used for the three-way classification problem.
Each case is plotted according to its values for $x_1$ and $x_2$, with the
plot symbol indicating the class, as follows: \protect\\[10pt]
  \hspace*{12pt} Class 0 = filled square \protect\\
  \hspace*{12pt} Class 1 = plus sign \protect\\
  \hspace*{12pt} Class 2 = open triangle}\label{fig-data}
}
\vspace*{45pt}

\mbox{~}

\includegraphics{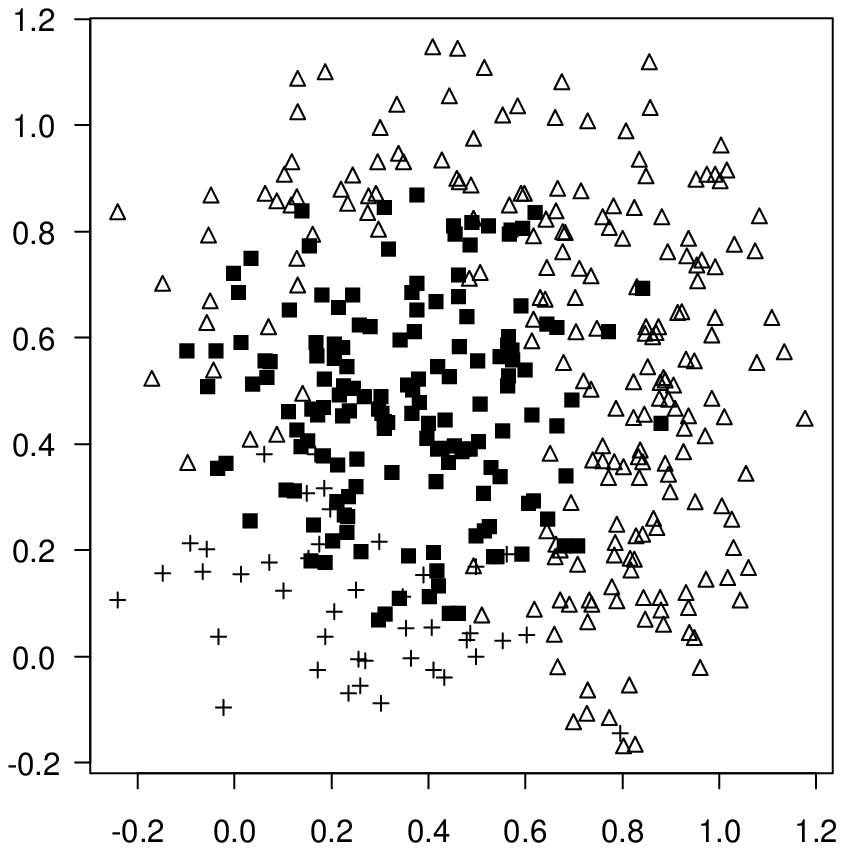}

\vspace*{20pt}

\end{figure}

To demonstrate the use of the software for classification, I applied it
to a synthetic three-way classification problem.  Pairs of data items,
$(x^{(i)}, t^{(i)})$ were generated by first randomly drawing
quantities $\tilde x^{(i)}_1$, $\tilde x^{(i)}_2$, $\tilde x^{(i)}_3$,
and $\tilde x^{(i)}_4$ independently from the uniform distribution
over the interval $(0,1)$.  The class of the item, $t^{(i)}$, encoded
as 0, 1, or 2, was then selected as follows:\ \ If the two-dimensional
Euclidean distance of $(\tilde x^{(i)}_1,\tilde x^{(i)}_2)$ from the
point $(0.4,0.5)$ was less than $0.35$, the class was set to 0;
otherwise, if $0.8*\tilde x^{(i)}_1+1.8*\tilde x^{(i)}_2$ was less
than $0.6$, the class was set to 1; and if neither of these conditions
held, the class was set to 2.  Note that $\tilde x^{(i)}_3$ and
$\tilde x^{(i)}_4$ have no effect on the class.  The inputs,
$x^{(i)}_1$, $x^{(i)}_2$, $x^{(i)}_3$, and $x^{(i)}_4$, available for
prediction of the target were the values of $\tilde x^{(i)}_1$,
$\tilde x^{(i)}_2$, $\tilde x^{(i)}_3$, and $\tilde x^{(i)}_4$ plus
independent Gaussian noise of standard deviation 0.1.  I generated
1000 cases in this way, of which 400 were used for training the model,
and 600 for testing the resulting predictive performance.  The 400
training case are shown in Figure~\ref{fig-data}.

This data was modeled using a Gaussian process for the latent values,
$y^{(i)}$, whose covariance function consisted of three terms --- a
constant part (fixed at 10), an exponential part in which the
magnitude, $\eta$, and the scales for the four inputs, $\rho_u$, were
variable hyperparameters, and a jitter part, fixed at $J=10$.  The fairly
large amount of jitter produces an effect close to a probit model, as
discussed in Section~\ref{sec-gp}.  Since each of the $\rho_u$ can
vary separately (under the control of a common higher-level
hyperparameter), the model is capable of discovering that some of the
inputs are in fact irrelevant to the task of predicting the target.
We hope that the posterior distribution of $\rho_u$ for these
irrelevant inputs will be concentrated near zero, so that they will
not degrade predictive performance.

The ``persistent'' form of hybrid Monte Carlo was used in the
sampling, as this allows the latent values to be resampled between
each leapfrog update of the hyperparameters.  A fairly low persistence
was used for the first few leapfrog updates, in order to allow energy
to be dissipated rapidly at first (through replacement of the
momentum, and consequent elimination of kinetic energy).  A larger
persistence was used thereafter, in order to suppress random walk
behaviour.  Before every update of the hyperparameters, the latent
values associated with training cases were updated using 100 Gibbs
sampling scans.  A sequence of five of these combined Gibbs sampling
and leapfrog updates were done in each sampling iteration, after which
the hyperparameters and latent values were saved for possible later
use.  Sampling was continued for 100 such iterations (500 leapfrog
updates), which took about 220 minutes on our SGI machine.

Complete details regarding the model and the sampling procedure used
may be found in the software documentation, where this problem is also
used as an example.

\begin{figure}[p]

\vspace*{3in}

\includegraphics{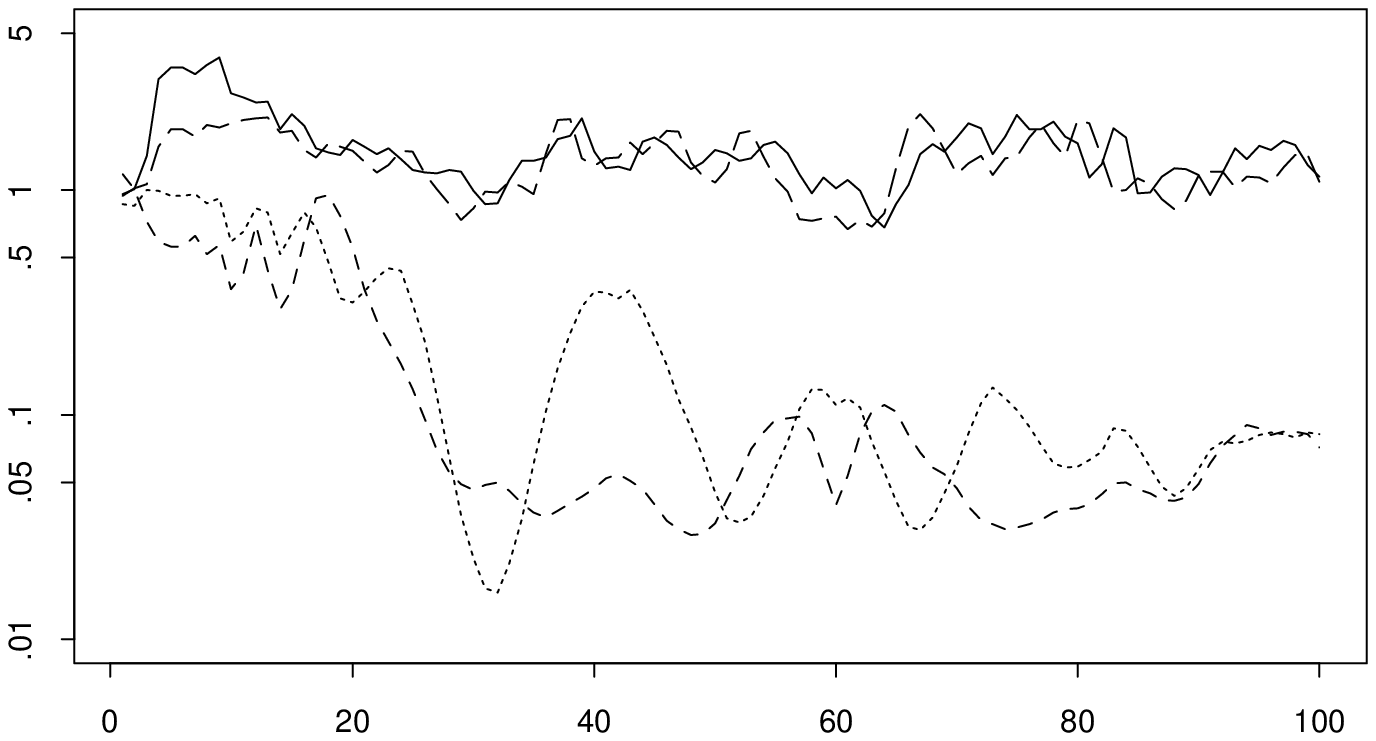}

\vspace*{-0.1in}

\caption[]{Progress of the four relevance hyperparameters during the
course of the Markov chain simulation.  The values are plotted on a
log scale, with $\rho_1$ = solid, $\rho_2$ = long dash, \mbox{$\rho_3$
= short dash,} and $\rho_4$ = dotted.}\label{fig-relevance}

\end{figure}

The convergence of the Markov chain simulation can be assessed by
plotting how the values of the hyperparameters change over the course
of the simulation.  Figure~\ref{fig-relevance} shows the progress of
the $\rho_u$ hyperparameters in the exponential part of the
covariance.  As hoped, we see that by about iteration 50, an apparent
equilibrium has been reached in which the hyperparameters $\rho_3$ and
$\rho_4$, associated with the irrelevant inputs, have values that are
much smaller than those for $\rho_1$ and $\rho_2$, which are
associated with the inputs that provide information about the target
class.

\begin{figure}[p]

\vspace*{3in}

\includegraphics{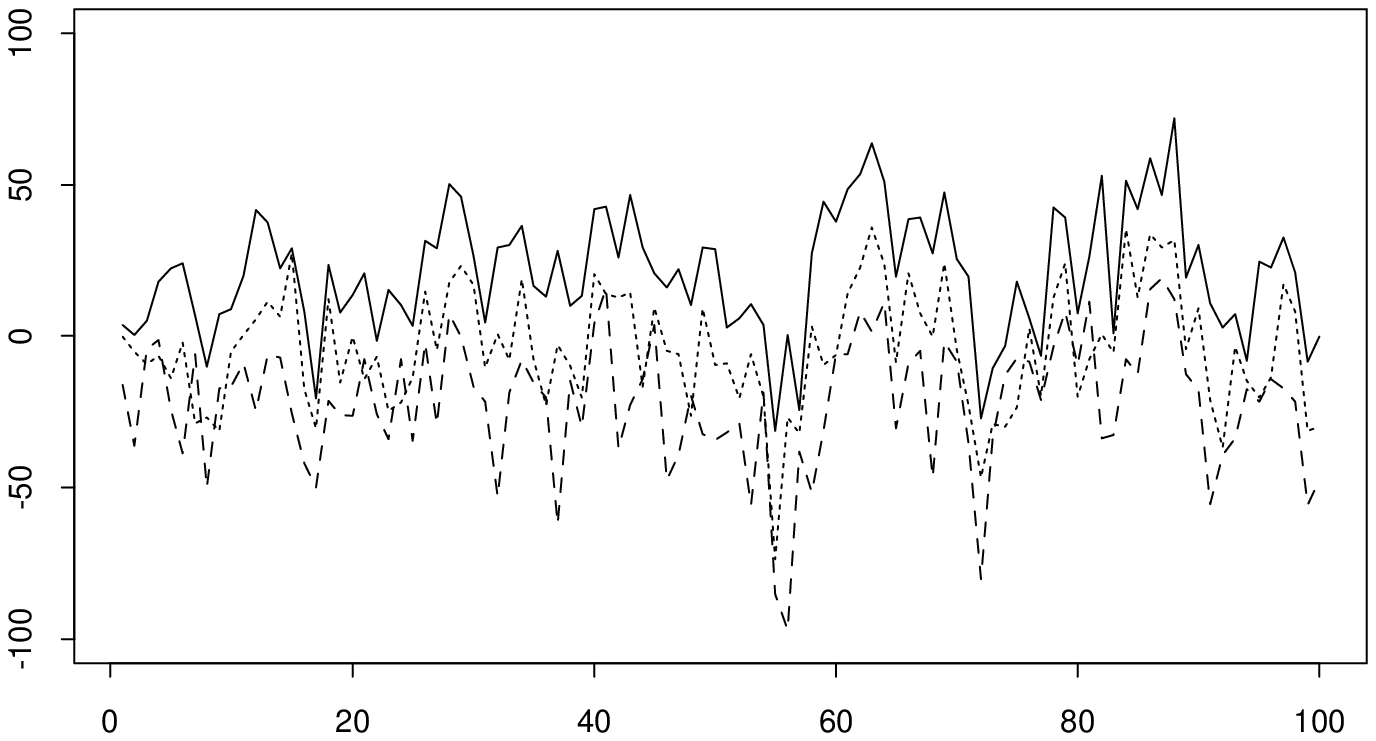}

\vspace*{-0.1in}

\caption[]{The latent values associated with one training case, for
which $x_1=0.20626$, $x_2=0.56059$, and $t=0$, over the course of the
Markov chain simulation.  The three latent values are shown as
Class 0 = solid, Class 1 = dashed, and Class 2 = dotted.}\label{fig-latent}

\end{figure}

The Markov chain simulation also updates the three latent values
associated with each training case, which define the class
probabilities by equation~(\ref{eq-softmax}).  These latent values for
a particular training case are plotted over the course of the
simulation in Figure~\ref{fig-latent}.  The Gibbs sampling scans
appear to be effective in moving these values about their equilibrium
distribution fairly rapidly.

To make predictions for test cases, we can average together the
predictive probabilities based on iterations after equilibrium was
apparently reached.  To reduce computation time, only every fifth
iteration was used, starting at iteration 55 (for a total of ten
iterations).  For each such iteration, the covariance matrix for the
latent values in training cases was inverted, after which the
predictive mean and variance for the latent values in each of the 600
test cases was found, using the latent values for training cases saved
for that iteration.  A sample of 100 points from this predictive
distribution was used to produce a Monte Carlo estimate of the
predictive probabilities for each of the three classes.  The final
predictive probabilities were found by averaging the predictions found
in this way for each of the iterations used.  The guess for the class
in a test case was the one with the largest predictive probability.

This procedure took about 11 minutes on our SGI machine.  The
classification error rate on the 600 test cases was 13\%.  This
performance is close to that of an analogous neural network model.  A
proper comparison of predictive performance with that of other
classification methods is beyond the scope of this paper. (Rasmussen
(1996) has done extensive comparisons of Gaussian process models with
other methods for regression problems.)

As expected, the time required for this problem varies considerably
with the number of training cases.  With only 100 training cases, the
time for the Markov chain simulation was about 16 minutes and the time
required to make predictions for the 600 test cases was less than a
minute.  The classification error rate using only the first 100
training cases was 17\%.

\section{Example:\ \ A regression problem with 
outliers}\label{sec-exo}\vspace*{-10pt}

To demonstrate how the software can be used to handle regression
problems with outliers, I applied a Gaussian process model with
non-Gaussian noise to a simple synthetic problem with a single input
variable.  Cases were generated in which the input variable, $x$, was
drawn from a standard Gaussian distribution, and the corresponding
target value came from a distribution with mean of 
\begin{eqnarray}
  0.3\ +\ 0.4\,x\ +\ 0.5\,\sin(2.7\,x)\ +\ 1.1\,/\,(1+x^2)
\end{eqnarray} 
For most cases, the distribution of the target about this mean was
Gaussian with standard deviation 0.1.  However, with probability 0.05,
a case was made an ``outlier'', for which the standard deviation was 
1.0 instead.

\begin{figure}[t]

\vspace*{3in}

\includegraphics{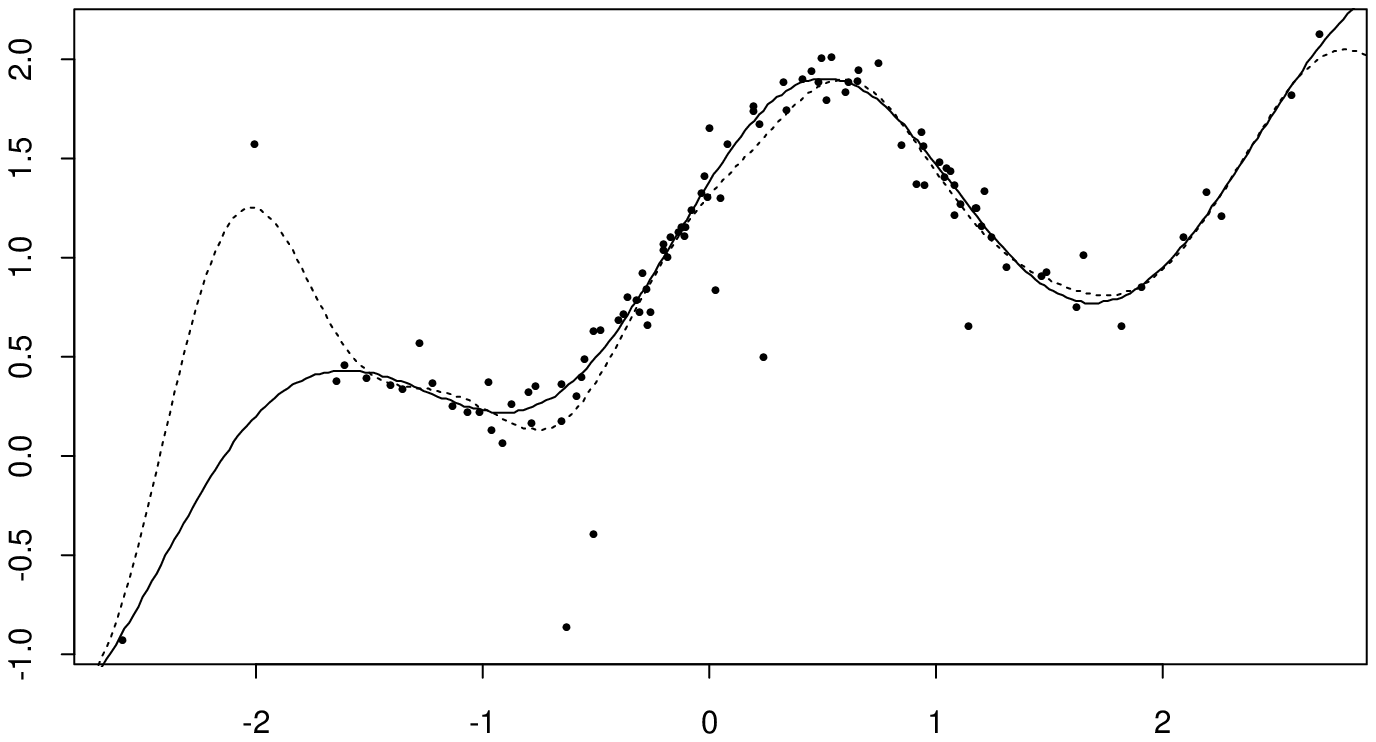}

\vspace*{-0.1in}

\caption[]{The regression problem with outliers.  The 100 training cases
are shown as dots, with the input on the horizontal axis, and the
target on the vertical axis.  The solid line gives the mean of the predictive
distribution using a model in which the noise was assumed to come from a 
$t$~distribution with 4~degrees of freedom.  The dotted line
gives the mean of the predictive distribution using a model in which 
the noise was assumed to be Gaussian.
}\label{fig-out}

\end{figure}

This data was modeled using a Gaussian process for the expected value
of the target, with the noise assumed to come from a $t$~distribution
with 4~degrees of freedom.  This is not particularly close to the
actual noise distribution, as described above, but the heavy tails of
the $t$~distribution may nevertheless allow this data to be modeled
without the outliers having an undue effect.  For comparison, the data
was also modeled under the assumption of Gaussian noise.  The data and
the predictions from these two models are shown in
Figure~\ref{fig-out}.

For these models, the covariance function used contained a constant
part (fixed at 1) and an exponential part (with variable
hyperparameters).  The model with non-Gaussian noise also included a
small amount of jitter ($J = 0.001$), in order to improve the
conditioning of the matrix computations used to sample for the latent
values.  This jitter is equivalent to a small amount of additional
noise in the model; since the amount of other noise is a variable
hyperparameter, the only real effect is to constrain the total noise
to be no less than the jitter.

Markov chain sampling for the model with $t$-distributed noise was
done by alternating hybrid Monte Carlo updates for the hyperparameters
(each consisting of 20 leapfrog updates) with updates for the
case-by-case noise variances.  Latent values were generated in order
to allow Gibbs sampling updates for the noise variances, using
equations~(\ref{eq-predm-g}) and~(\ref{eq-predv-g}), but were
discarded thereafter.  The Markov chain was simulated for 200 such
iterations, and predictions were then made based on every fifth
iteration after iteration 100.  The time required for the simulation
was about six minutes on our SGI machine.

Further details of the model and the Markov chain method can be
obtained from the description of this example in the software
documentation.

As can be seen in Figure~\ref{fig-out}, the model with $t$-distributed
noise produces predictions that seem more reasonable than those
produced by the model with Gaussian noise, based just on looking at
the scatterplot of the data.  The predictions using $t$-distributed
noise are also closer to the true function.

\section{Discussion}\label{sec-disc}\vspace*{-10pt}

The software described in this paper extends the scope of Gaussian
process models to classification problems and to regression problems
with non-Gaussian noise, by using a Markov chain Monte Carlo method in
which latent values for each case are represented.  Models can be
based on a variety of covariance functions, which can be defined in
terms of hyperparameters with hierarchical priors.  The implementation
also allows a wide variety of Markov chain sampling methods to be
used.

With these facilities, the usefulness of Gaussian process models for a
variety of problems can be explored.  The examples in this paper show
that Gaussian process models can be practically applied to
classification problems of moderate size, and to regression problems
with non-Gaussian noise; other examples of regression and
classification models are included in the software documentation.  One
major focus for future work is to explore the uses of elaborate
covariance functions in real problems.  The fairly simple model of
Section~\ref{sec-exc} illustrates how the hyperparameters defining the
covariance function can adaptively determine how relevant the various
inputs are for predicting the target.  The range of covariance
functions implemented permits hierarchical models that are more
elaborate than this.  For example, by including several exponential
parts in the covariance function, each with a separate set of
relevance hyperparameters, it is possible to define a prior
distribution that puts considerable prior weight on models that are of
a nearly additive form, in which the function is decomposed into the
sum of several functions, each of which depends on only a small subset
of the inputs.  Such a model can automatically determine an
appropriate additive decomposition, if an additive model is in fact
appropriate.  This mirrors a similar idea for neural network models
(Neal 1996, Section 5.2).

The implementation described here is rather straightforward.  Most
operations are performed in the simplest way that gives acceptable
results.  A number of modifications can be contemplated.  Faster
convergence could probably be obtained by updating the latent
variables using hybrid Monte Carlo rather than Gibbs sampling.
Computation time for matrix operations might be reduced by using the
conjugate gradient approach of Gibbs and MacKay (1997a).  In another
direction, one might look for ways of reducing or eliminating the need
for jitter in the covariance function, since although this appears to
usually be an acceptable solution to the problem of poorly conditioned
matrices, there may be some circumstances where it is undesirable,
such as when using a Gaussian process to model noise-free data from
``computer experiments'' (eg, Sack, Welch, Mitchell, and Wynn 1989).

Even without further algorithmic improvements, Gaussian process models
are now feasible for datasets of up to about a thousand cases, using
fairly run-of-the-mill computers, provided one is willing to wait up
to several hours for results on the larger datasets.  Using these
models is therefore a feasible option for many regression and
classification problems.  Despite the fairly unfavourable $n^2$ growth
in memory requirements and $n^3$ growth in computation time of the
present Gaussian process algorithms, improvements in computer
technology over the next few years will likely allow these models to
be applied to most problems encountered in practice.  Because of the
ease with which flexible hierarchical models can be defined using
Gaussian processes, I believe that they will prove to be among the
most useful techniques for nonparametric regression and
classification.

\section*{Acknowledgements}\vspace*{-10pt}

I thank Carl Rasmussen for many helpful discussions, and for the
opportunity to learn from the implementation of Gaussian process
regression that he used for his thesis.  I also thank David MacKay and
Chris Williams for their comments on the manuscript.  This research
was supported by the Natural Sciences and Engineering Research Council
of Canada.

\section*{References}\vspace*{-10pt}

\leftmargini 0.2in
\labelsep 0in

\begin{description}
\itemsep 2pt

\item
  Barber, D.\ and Williams, C.~K.~I.\ (1997) ``Gaussian processes
  for Bayesian classification via hybrid Monte Carlo'', to appear
  in \textit{Advances in Neural Information Processing Systems~9}.

\item
  Duane, S., Kennedy, A.~D., Pendleton, B.~J., and Roweth, D.\ (1987)
  ``Hybrid Monte Carlo'', \textit{Physics Letters B}, vol.~195, pp.~216-222.

\item 
  \mbox{Gibbs, M.~N.\ and MacKay, D.~J.~C.\ \ (1997a)}
  ``Efficient implementation of Gaussian processes'', draft manuscript.

\item 
  \mbox{Gibbs, M.~N.\ and MacKay, D.~J.~C.\ \ (1997b)}
  ``Variational Gaussian process classifiers'', draft manu\-script.

\item
  Gilks, W.~R.\ and Wild, P.\ (1992) ``Adaptive rejection sampling
  for Gibbs sampling'', \textit{Applied Statistics}, vol.~41, pp.~337-348.

\item
  Horn, R.~A.\ and Johnson, C.~R.\ (1985) \textit{Matrix Analysis},
  Cambridge University Press.

\item
  Horowitz, A.~M.\ (1991) ``A generalized guided Monte Carlo algorithm'',
  \textit{Physics Letters B}, vol.~268, pp.~247-252.

\item
  Mardia, K.~V.\ and Marshall, R.~J.\ (1984) ``Maximum likelihood estimation
  of models for residual covariance in spatial regression'', 
  \textit{Biometrika}, vol.~71, pp.~135-146.

\item
  Neal, R.~M.\ (1993) \textit{Probabilistic Inference Using Markov Chain
  Monte Carlo Methods}, Technical Report CRG-TR-93-1, Department
  of Computer Science, University of Toronto.  Available in Postscript
  via URL {\tt http://www.cs.utoronto.ca/$\sim$radford/}.

\item
  Neal, R.~M.\ (1994) ``An improved acceptance procedure for the 
  hybrid Monte Carlo algorithm'', \textit{Journal of Computational Physics},
  vol.~111, pp.~194-203.

\item
  Neal, R.~M.\ (1996) \textit{Bayesian Learning for Neural Networks},
  New York: Springer-Verlag.

\item
  O'Hagan, A. (1978) ``Curve fitting and optimal design for prediction''
  (with discussion), \textit{Journal of the Royal Statistical Society~B}, 
  vol.~40, pp.~1-42.

\item
  O'Hagan, A.\ (1994) \textit{Bayesian Inference} (Volume 2B in 
  Kendall's Advanced Theory of Statistics). 

\item 
  Rasmussen, C.\ (1996) \textit{Evaluation of Gaussian Processes and
  other Methods for Non-Linear Regression}, Ph.D.\ Thesis, University of
  Toronto, Department of Computer Science. Available in Postscript
  via URL {\tt http://www.cs.utoronto.ca/$\sim$carl/}.

\item
  Sacks, J., Welch, W.~J., Mitchell, T.~J., and Wynn, P.\ (1989)
  ``Design and analysis of computer experiments'' (with discussion), 
  \textit{Statistical Science}, vol.~4, pp.~409-435.

\item
  Thisted, R.~A.\ (1988) {\em Elements of Statistical Computing},
  New York: Chapman and Hall.

\item
  von Mises, R.\ (1964) \textit{Mathematical Theory of Probability and
  Statistics}, New York: Academic Press.

\item
  Wahba, G.\ (1978) ``Improper priors, spline smoothing and the 
  problem of guarding against model errors in regression'', \textit{Journal
  of the Royal Statistical Society~B}, vol.~40, pp.~364-372.

\item
  Williams, C.~K.~I.\ and Rasmussen, C.~E.\ (1996) ``Gaussian processes
  for regression'', in D.~S.~Touretzky, M.~C.~Mozer, and M.~E.~Hasselmo 
  (editors) \textit{Advances in Neural Information Processing Systems~8}, 
  MIT Press.

\item
  Yaglom, A.~M.\ (1987) \textit{Correlation Theory of Stationary and
  Related Random Functions, Volume I: Basic Results}, New York: Springer-Verlag.

\end{description}
\end{document}